\documentclass[aps,prd,twocolumn,showpacs,groupedaddress,nofootinbib]{revtex4}
\usepackage{graphicx}
\begin{document}

\title{Revisiting Glueball Wave Functions at Zero and Finite Temperature}

\author{Mushtaq Loan}
\email{mushe@phys.unsw.edu.au}
\affiliation{International School,
Jinan University, Guangzhou 510632, China\\
School of Physics, The University of
New South Wales, Sydney, NSW 2052, Australia}

\date{\today}

\begin{abstract}
We study the sizes and thermal properties of glueballs in a three
dimensional compact Abelian gauge model on improved lattice. We
predict the radii of $\sim 0.60$ and $\sim 1.12$ in the units of
string tension, or $\sim 0.28$ and $\sim 0.52$ fm, for the scalar
and tensor glueballs, respectively. We perform a well controlled
extrapolation of the radii to the continuum limit and observe that
our results agree with the predicted values. Using Monte Carlo
simulations, we extract the pole-mass of the lowest scalar and
tensor glueballs from the temporal correlators at finite
temperature. We see a clear evidence of the deconfined phase, and
the transition appears to be similar to that of the
two-dimensional $XY$ model as expected from universality
arguments. Our results show no significant changes in the glueball
wave functions and masses in the deconfined phase.
\end{abstract}

 \maketitle

\section{INTRODUCTION}
The prediction of glueball masses has long been attempted in
lattice gauge theory calculations
\cite{Morningstar99,Vaccarino99,Teper98,Norman98,Bali93}. These
calculations show that the lowest-lying scalar, tensor and axial
vector glueballs lie in the mass region of 1-2.5 GeV. While there
is a long history of glueball mass calculations in lattice QCD,
little is known about the glueballs besides their masses. Accurate
lattice calculations of their size, matrix elements and form
factors would help considerably in their experimental
identification.

Glueball wave functions and sizes have been studied in the past
\cite{DeGrand87,Forcrand92,Boyd96,Ishii02}, but much of the early
work contains uncontrolled systematic errors, most notably from
discretisation effects. The scalar glueball is particularly
susceptible to such errors for the Wilson gauge action, due to the
presence of a critical end point of a line of phase transitions in
the fundamental-adjoint coupling plane. As this critical end-point
(which defines the continuum limit of a $\phi^{4}$ scalar field
theory) is neared, the coherence length in the scalar channel
becomes large, which means that the mass gap in this channel
becomes small; glueballs in other channels seem to be affected
very little. Results in which the scalar glueball was found to be
significantly smaller than the tensor were most likely due to
contamination of the scalar glueball from this non-QCD critical
point \cite{Forcrand92}. On the other hand, the calculations using
operator overlaps obtained from variational optimization for
improved lattice gauge action, which are designed to avoid
spurious critical point, show that the scalar and tensor glueballs
were of typical hadronic dimensions \cite{Morningstar99,Ishii02}.
A straightforward procedure to address the controversy over
glueball size is to measure the glueball wave function, much
 in the same way as the meson and baryon wave functions
were measured \cite{Alexandrou02}.

In this paper, we study the low-lying scalar and tensor glueballs
and their wave functions with renormalized tadpole improved
Symanzik gauge action \cite{Sakai02}. Our techniques for
calculating the glueball wave functions from Wilson loop operators
are outlined in Sec. \ref{sec2}. We present and discuss our
results at zero temperature in Sec. \ref{sec3}. We extend our
method to examine the wave functions and masses at finite
temperature in this section. Here we give an explicit
interpretation of deconfinement in terms of the power-law
behaviour of the correlation function. Section \ref{sec4} is
devoted to the summary and concluding remarks.

\section{Wave functions of glueballs}
\label{sec2}
In contrast with the techniques used in Ref.
\cite{Forcrand92}, we measure our lattice operators from spatially
connected Wilson loops. Glueballs are colour singlet states and
one should be able to construct them with closed-loop paths which
are gauge invariant. The choice of such loops eliminate the need
for gauge fixing\footnote{It should be noted that gauge-invariant
Wilson loops have $a^{4}$ dependence compared to $a^{2}$
dependence of two-link operator used in Ref.
\protect\cite{Forcrand92}. The lower dimension operator yields a
linear dependence in the correlation function as opposed to
$a^{5}$ dependence for Wilson loops. This improves the glueball
signal as $a$ is reduced.}. Although
the calculations
in Ref. \cite{Forcrand92} have produced some interesting results,
the approach suffers from a basic problem: the observables are
calculated from a lattice version of the 2-glue operator, which
risks a mixture of glueball states with flux states\footnote{The
link-link operator used in Ref. \protect\cite{Forcrand92} sums up
a large number of loops; some of these loops have a zero winding
number and project on glueballs - others have non-zero winding
number and project on flux states also called torelons.}.

In this study we take a more direct approach to the problem. We
measure the observables in a three-step procedure. First, we
calculate the lattice operator
\begin{equation}
\Phi (\vec{r},t)=\sum_{\bf x}\left[\phi (\vec{x},t)+\phi
(\vec{x}+\vec{r},t)\right],
\label{eqn01}
\end{equation}
where $\phi$ is the plaquette operator and $\Phi$ measures the two
plaquette or two-loop component of the glueball wave function. The
$r$ dependence will be reflected in the length of links required
to close the loops. From a suitable linear combinations of
rotation, parity inversions and real or imaginary parts of the
operators involved in $\Phi$, one can construct glueball operators
with desired quantum numbers \cite{Teper98,loan03,Mushe04}. Since
we want to explore the nature of wave functions, we focus only on
the low-lying ``symmetric'' and ``antisymmetric'' scalar channels
(which are the cosine and sine, respectively of the Wilson loop in
question) and tensor glueball states.

The wave function and mass are obtained from the correlation
function:
\begin{equation}
C(\vec{r},t)= \langle
\Phi^{\dagger}(\vec{r},t)\Phi(0,0)\rangle ,
\label{eqn02}
\end{equation}
where one needs to subtract the vacuum contribution from the
correlator for $0^{++}$ state. The source can be held fixed while
the sink takes on the r dependence. This proves to be helpful in
maintaining a good signal. The disentangling of the glueball and
torelon is usually taken care of automatically by the choice of
Wilson or Polyakov loops.

To increase the overlap with the lowest state and reduce the
contamination from higher states, we exploit the APE link smearing
techniques \cite{APE87}. The procedure is implemented by an
iterative replacement of the original spatial link variable by a
smeared link. This results in correlations which reach their
asymptotic behaviour at small time separations. In addition, the
noise from ultraviolet fluctuations is reduced. The smearing
parameter is fixed to $0.7$ and ten iterations of the smearing
process are used. To find the optimum smearing value, $n$, we
examine the ratio (at $r=0$ and 1)
\begin{displaymath}
C(r,t+1)/C(r,t),
\end{displaymath}
which should be maximum for good ground state dominance. Using
$1\times 1 $ loop as template, the best signal is obtained with
four smearing steps, with $1\times 1$ and $2\times 2$ loops being
almost indistinguishable.  At $\beta =2.0$, the signal in $1\times
1$ showed a slow convergence with $n$, hence $2\times 2$ loops
were preferred for optimum overlap. A typical value which proved
to be sufficient for this case was $n=2$.

A second pass was made to measure the optimized correlation matrices
\begin{equation}
C_{ij}(t)=\langle \Phi (r_{i},t)\Phi (r_{j},0)\rangle
- \langle\Phi(r_{i})\rangle\langle\Phi(r_{j})\rangle .
\label{eqn03}
\end{equation}

Let $\psi^{(k)}$ be the radial wave function of the $k$-th
eigenstate of the transfer matrix, then
\begin{equation}
C_{ij}(t)=\sum_{k}\alpha_{k}\psi^{(k)}(r_{i})\psi^{(k)}(r_{j})\mbox{e}^{-m_{k}t}.
\label{eqn04}
\end{equation}
The glueball masses and the wave functions are extracted from the
Monte Carlo average of $C_{ij}(t)$ by diagonalizing the
correlation matrices $C(t)$ for successive times $t$:
\begin{equation}
C(t)=\tilde{R}(t)D(t)R(t),
\label{eqn05}
\end{equation}
where $D$ is a diagonal matrix of the eigenvalues and $R$ a
rotation matrix whose columns are the eigenvectors of $C$. Each
eigenvector of $C$ matches an eigenstate $\psi^{(k)}(r)$ of the
complete transfer matrix. As the wave function is largest at the
origin, one would first determine the glueball mass with the
optimal separation, and then fix that mass for all r, and extract
the wave function for less optimal separations. Similar to the case of
mesons \cite{Velikson85}, the wave function is expected to decrease
exponentially with the $r$ at large separations  and is therefore fitted
with the
simple form
\begin{equation}
\psi (r) \equiv \mbox{e}^{-r/r_{0}}
\label{eqn06}
\end{equation}
to determine the effective radius $r_{0}$.
The effective mass can be read off directly from the largest
eigenvalue corresponding to the lowest energy
\begin{equation}
m_{eff} =
\mbox{log}\left[\frac{\lambda_{0}(r=0,t=1)}{\lambda_{0}(r=0,t=2)}
\right]
\label{eqn07}
\end{equation}

\section{Simulation results and discussion}
\label{sec3}
\subsection{Results at zero temperature}
\label{subsec3-1}

Most of our Monte Carlo calculations are carried out on
$16^{2}\times 16$ lattice with periodic boundary conditions
($16^{2}$ is the space-like box and $16$ is the extension in
Euclidean time direction). The gauge configurations are generated
using the Metropolis algorithm. After the equilibration,
configurations are stored every 250 sweeps; 3000 stored
configurations are used in the measurement of glueball masses.
Measurements made on the stored configurations are binned into 10
blocks with each block containing an average of 300 measurements.
The mean and the final errors are obtained using
single-elimination jackknife method with each bin regarded as an
independent data point. Three sets of measurements were taken at
$\beta = 2.0, 2.25$ and $2.5$. Some finite-size consistency checks
are done at $\beta =2.25$ on an $20^{2}\times 20$ lattice.

\begin{figure}[!h]
\scalebox{0.45}{\includegraphics{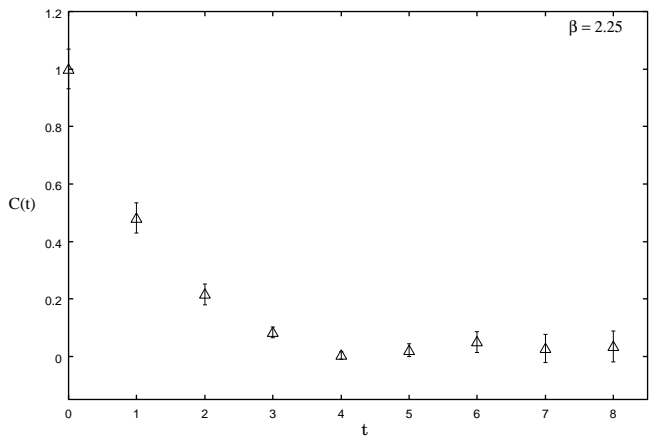}}
\caption{
\label{figcorr}
Correlation function for the $0^{++}$ channel against $t$.}
\end{figure}

The glueball correlation function
for the $0^{++}$ channel against $t$ at $\beta =2.25$ is shown in
Fig. \ref{figcorr}. It can be seen that the expected behaviour of
correlation  function is attained
virtually straight away. The absolute errors in the correlation
functions are expected to be independent of $t$ for large $t$.
Our errors are consistent with this expectation.

\begin{figure}[!h]
\scalebox{0.45}{\includegraphics{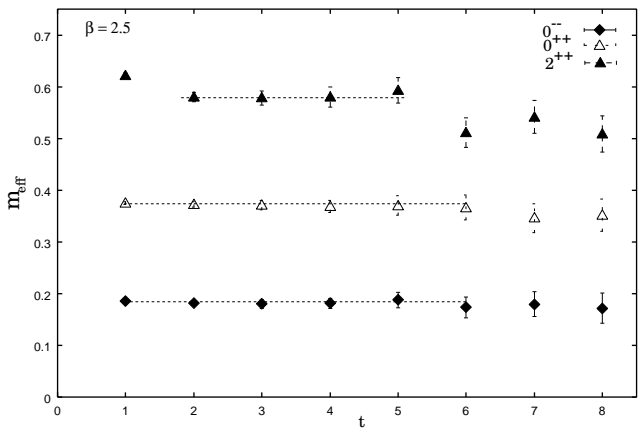}} \caption{
\label{figeffmass}
Effective mass plot for scalar and tensor glueball states for $\beta =2.5$.
The dashed horizontal lines indicate the plateau values.}
\end{figure}

Effective mass plot for $\beta= 2.5$ simulation is presented in
Fig. \ref{figeffmass}. For $0^{++}$ and $0^{--}$ channels each  it
was possible to find a fit region $t_{\mbox{min}} -
t_{\mbox{max}}$ in which convincing plateaus were observed. The
effective masses are found stable using different values of t in
Eq. (\ref{eqn07}), which suggests that the glueball ground state
is correctly projected. At $\beta =2.5$, we noticed considerable
fluctuations in the tensor mass at large $t$. An acceptable fit
was only possible for $t =[2 - 5]$. To ensure the validity of our
results, we compared them to those obtained using
\begin{equation}
m'_{eff}=\mbox{log}
\left[\frac{\lambda_{0}(t-1)-\lambda_{0}(t)}{\lambda_{0}(t)-\lambda_{0}(t+1)}
\right].
 \label{eqn07b}
\end{equation}
It was found that the evaluations of Eqs. (\ref{eqn07}) and
(\ref{eqn07b}) yielded very consistent within statistical errors.

\begin{figure}[!h]
\scalebox{0.45}{\includegraphics{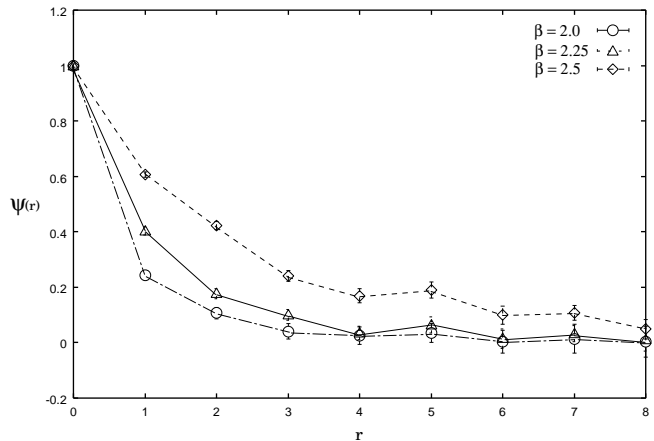}}
\caption{
\label{fig0pp}
Scalar $0^{++}$ glueball wave functions measured on a $16^{3}$ lattice for
various values of $\beta$.}
\end{figure}

\begin{figure}
\scalebox{0.45}{\includegraphics{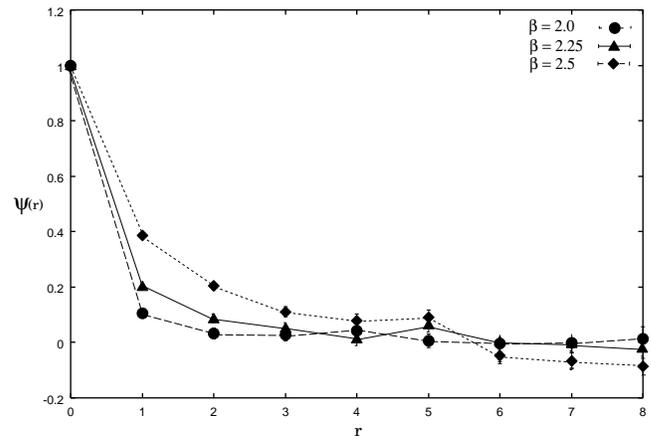}} \caption{
\label{fig0mm} Scalar $0^{--}$ glueball wave function on a
$16^{3}$ lattice at $\beta =2.0, 2.25$ and 2.5.}
\end{figure}

\begin{figure}
\scalebox{0.45}{\includegraphics{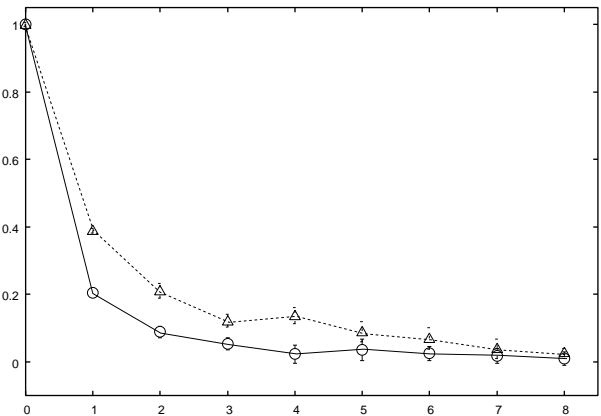}} \caption{
\label{fig0mm20} Scalar $0^{--}$ glueball wave function on a
$20^{3}$ lattice at $\beta =2.25$ (open circles) and $\beta =2.5$
(open triangles).}
\end{figure}

The Wave functions are extracted at time-separations $t=1$ and 2.
We found a little variation (less than two percent) in the eigenvectors of
$C(t)$ with $t$
which suggests that there is no mixing with states of
distinct masses. Typical plots of the wave functions,
normalised to unity at the
origin, for the symmetric and antisymmetric scalar glueballs, at
$\beta = 2.0, 2.25$ and 2.5 are shown in Figs. \ref{fig0pp} and
\ref{fig0mm} respectively. For guiding the eyes the Monte Carlo
points of the same $\beta$-value are connected with straight
lines. The scalar wave function shows the expected behaviour for
all the $\beta$ values analysed here.
As for the antisymmetric channel we notice the presence of
negative contributions in the glueball wave function for $r>6$ at
$\beta=2.5$. However these contributions do not persist when the
lattice size is increased from $L=16$ to 20 (Fig. \ref{fig0mm20}).
This would mean that these effects are unphysical and can be
described as a finite volume artifact.

For this reason we extract the effective radius of  the
antisymmetric state from the wave function obtained at larger
volume\footnote{The results for $16^{3}$ lattice in Fig.
\ref{fig0mm} are shown only as an illustration. Comparison of the
data for the effective mass on two lattice sizes reveals that none
of our states could be interpreted as a torelon pairs, since no
mass reduction of sufficient magnitude was found as the lattice
volume was reduced.}. The symmetric scalar glueball wave function,
on the other hand, barely changes sign.

\begin{figure}
\scalebox{0.45}{\includegraphics{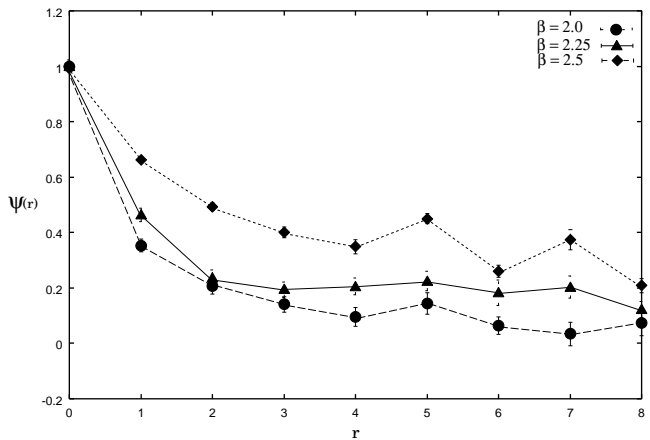}} \caption{
\label{fig2pp} Tensor  glueball wave function on a $16^{3}$ lattice
at $\beta =2.0, 2.25$ and 2.5.}
\end{figure}

Fig. \ref{fig2pp} shows the wave function for the tensor glueball,
at $\beta = 2.0, 2.25 $ and 2.5. The tensor wave function remains positive
and shows the expected flatness. It can be seen that tensor
glueball is much more extended than the scalar as one moves
towards higher $\beta$ values. This would imply that tensor is
therefore more sensitive to the finite-size effects, which is very
visible in the distortion of the wave function for large $r$ at
$\beta = 2.5$. Naively we would expect that the spatial size at which we
begin to encounter large finite size effects to be related to the size of
the glueball.

The expected finite-size scaling behaviour of the mass gap near
the continuum critical point in this model is not known; but
Weigel and Janke \cite{Weigel99} have performed a Monte Carlo
simulation for an O(2) spin model in three dimensions which should
lie in the same universality class, obtaining
\begin{equation}
M\approx 1.3218/L
\label{eqn08}
\end{equation}
for the magnetic gap. In order to ascertain the finite-size effect
on our measurements, we performed extra simulations on a
$20^{2}\times 20$ lattice at $\beta =2.25$ and 2.5. The  mass
and size of $0^{++}$ channel are almost unchanged as the lattice
size increases from 16 to 20. We also find that our estimates for
the tensor state are consistent with no finite volume dependence
at $\beta = 2.25$. However, the tensor mass was found to increase
by about $4\%$ and the effective radius by about $7\%$ from 16 to 20
lattices at $\beta = 2.5$. We do not
have enough data extrapolate mass and the radius to the infinite
volume limit or to check whether the difference is due to statistical
errors or whether there is an incomplete convergence.
Given that no mass reductions of sufficient magnitudes were
found as the lattice volume is changed, none of our states could be
interpreted as a torelon pair.

\begin{figure}
\scalebox{0.45}{\includegraphics{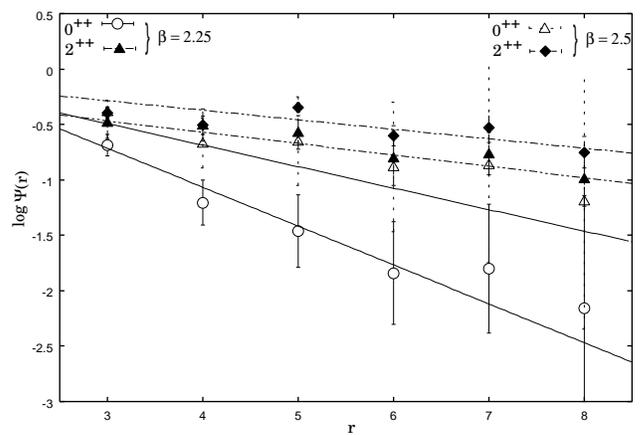}} \caption{
\label{figlogwf} The logarithmic plot of the scalar and tensor
glueball wave functions at $\beta = 2.25$ and 2.5. The effective
radii can be obtained from the inverse slopes of the curves.}
\end{figure}

In order to get some quantitative information on the  effective
radius, the glueball wave functions are fitted in the
range $3\leq r\leq 8$  by the form (\ref{eqn06}). This form fits the data
rather well for the scalar glueball with the best-fit estimates
obtained with a $\chi^{2}/N_{DF}$ of 0.92 - 0.67. Due to
distortion\footnote{Because of the distortion and impossible complete
elimination of all the excited-states, especially near $r\sim 0$, it
follows that Eq. (\protect\ref{eqn06}) holds only in the limited interval,
which does not include the vicinity of $r\sim 0$.} in the tensor
wave function at small $r$ at $\beta=2.5$, a
meaningful fit was possible only in the range $6\leq r\leq 8$. The
effective radius  obtained was confirmed by examining the
plateau in the ratio $\mbox{log}[\psi (r)/\psi (r+1)]$. Note, that our
lograthmically plotted wave functions (Fig. \ref{figlogwf}) are merely
illustrations.

To summarize: in the weak coupling region a spectrum of massive
$0^{++}$, $0^{--}$ and $2^{++}$ glueballs is indicated with
\begin{displaymath}
m(0^{--}) < m(0^{++}) < m(2^{++}).
\end{displaymath}
Since there is a  good signal for wave function persisting long
enough to demonstrate convergence to the asymptotic value, it
seems to be reasonable to estimate mass ratios with our present
method. The estimates of masses and $r_{0}$, in  lattice units, at
various $\beta$ values are shown in Tables \ref{tab1}, \ref{tab2}
and \ref{tab3}.
\begin{table}[!h]
\caption{ \label{tab1} Masses of scalar  glueballs in lattice
units for two spatial extensions, $L=16$ and 20.}
\begin{ruledtabular}
\begin{tabular}{cccccc}
   & &  \multicolumn{2}{c}{Mass}\\
    & \multicolumn{2}{c}{$0^{++}$}& \multicolumn{2}{c}{$0^{--}$} \\
 $\beta$/L & 16 & 20 &16  & 20    \\ \hline
2.0  & 0.803(6) &  & 0.441(4)&  \\
2.25 & 0.523(3) & 0.527(3) & 0.266(3)& 0.261(4) \\
2.5  & 0.364(3) & 0.369(2) & & 0.182(2) \\
\end{tabular}
\end{ruledtabular}
\end{table}

\begin{table}[!h]
\caption{ \label{tab2} Sizes of scalar glueballs in lattice units
for two spatial extensions, $L=16$ and 20.}
\begin{ruledtabular}
\begin{tabular}{cccccc}
   & &   \multicolumn{2}{c}{Size}\\
   & \multicolumn{2}{c}{$0^{++}$}& \multicolumn{2}{c}{$0^{--}$} \\
 $\beta$/L & 16 & 20 &16  & 20   \\ \hline
2.0   & 1.4(2)& & 1.0(1)& \\
2.25  & 2.75(4)& 2.8(3)& 2.14(4)& 2.2(2)\\
2.5   & 5.1(1.0)& 5.2(9)& & 5.0(1.0)\\
\end{tabular}
\end{ruledtabular}
\end{table}

\begin{table}[!h]
\caption{
\label{tab3}
Mass and size of tensor glueballs in
lattice units for two spatial extensions, $L=16$ and 20.}
\begin{ruledtabular}
\begin{tabular}{ccccc}
   &  \multicolumn{2}{c}{Mass} & \multicolumn{2}{c}{Size}\\
 $\beta$/L & 16 & 20 &16 & 20 \\ \hline
2.0 & 1.2(1) & & 5.0(7)& \\
2.25 & 0.82(2) & 0.81(6)& 9.7(1.7)& 9.8(1.4)\\
2.5 & 0.544(2) & 0.58(1)& 10.1(2.6)& 10.2(2.4)\\
\end{tabular}
\end{ruledtabular}
\end{table}
Our results for lattice masses and mass ratios are generally,
within statistical errors, in  agreement with the existing
Euclidean estimates \cite{Mushe03,Mushe04,Teper98}, if perhaps a
little high in places. Qualitatively our results, at zero
temperature, are in agreement with the scenario of spectrum of
massive magnetic monopoles.

\begin{table}[!h]
\caption{
\label{tab3b}
Glueball sizes in the units of string tension.}
\begin{ruledtabular}
\begin{tabular}{cccccc}
$\beta$ &$K (=a^{2}\sigma )$ &$a_{eff}$ &  $r_{0^{++}}\sqrt{\sigma}$ &
$r_{0^{--}}\sqrt{\sigma}$
& $r_{2^{++}}\sqrt{\sigma}$ \\ \hline
2.0  & 0.0508(5) & 0.0856 &  0.31(14) & 0.24(9) & 1.13(18) \\
2.25 & 0.0221(3) & 0.0481 &  0.40(17) & 0.32(16) & 1.14(22) \\
2.5  & 0.0119 & 0.0272 &  0.56(21) & 0.50(19) & 1.11(25)\\
\end{tabular}
\end{ruledtabular}
\end{table}

\begin{figure}[!h]
\scalebox{0.45}{\includegraphics{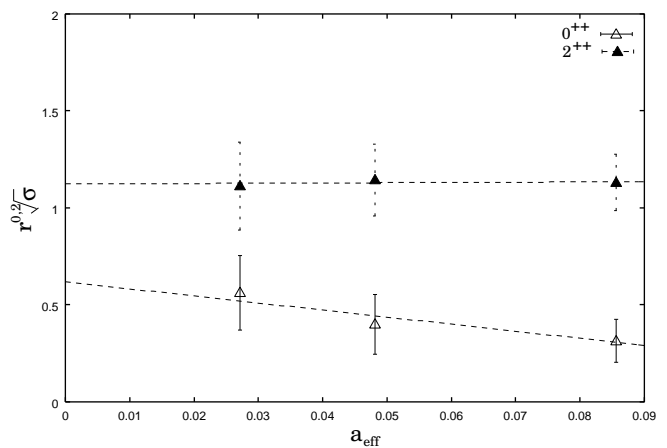}}
\caption{
\label{figcontrad}
Glueball radii in the units of string tension as a function of the
effective spacing, $a_{eff}$. Extrapolations to the continuum limit are
shown as dashed lines.}
\end{figure}

To extrapolate our  effective radii  to the continuum limit, we
take the dimensionless products of sizes so that the scale, $a$,
in which they are expressed cancels. We choose to take products of
the effective radii, $r_{0,2}/a$, to $a\sqrt{\sigma}$ since the
string tension is our most accurately calculated quantity. As in
the (3+1)D confining theories, we expect that dimensionless
product of physical quantities, such as $r_{0,2}\sqrt{\sigma}$,
will approach their continuum limit with correction of
$O(a_{eff}^{2})$, where $a_{eff}$ is the effective lattice spacing
in ``physical units" when the mass gap has been renormalized to a
constant \cite{Mushe03}. The string tension, $K (=a^{2}\sigma )$,
is obtained by using the Wilson loop averages and fitting the
on-axis data with $V(r)$. In Fig. \ref{figcontrad} we show the
product $r_{0,2}\sqrt{\sigma}$ plotted against $a_{eff}$. Since
the products are plotted against $a_{eff}$, the continuum
extrapolations are simple straight lines. We notice that the
product $r_{0,2}\sqrt{\sigma}$   varies only slightly over the
fitting range. The  non-zero lattice spacing values of the product
are within 0.04 - 0.29 and 0.01 - 0.02  standard deviations of the
extrapolated zero lattice spacing results for the scalar and
tensor glueballs respectively. The striking feature of this plot
is the little variation of the product with $a_{eff}$. This will
make for very accurate and reliable continuum  extrapolations.
Linear extrapolations to the continuum limit yield values of
$0.60\pm 0.05$ and $1.12\pm 0.03$, in the units of string tension,
for the scalar and tensor states, respectively. In contrast to the
tensor, the scalar glueball size shows significant finite-spacing
errors. By setting the string tension to 420 MeV, we obtain the
physical radii of 0.28(7)  and  0.52(5) fm, for the scalar and
tensor glueballs, respectively. Our results show  the size of the
tensor glueball roughly two  times as large as the scalar
glueball. These estimates agree with the rough estimates of
glueball sizes obtained at various temperatures in Ref.
\cite{Ishii02}. This is an improvement over the estimates obtained
in Ref. \cite{Forcrand92} where the predicted radius for the
tensor  glueball ($\sim 0.8$ fm) was found four times larger than
scalar glueball radius.

\subsection{Finite temperature results}
\label{subsec3-2}

To check the consistency of our method, we performed  a study on
an asymmetric lattice: $16^{2}\times 4$  at $\beta = 2.25$. The
procedure itself is a straight forward extension of the procedure
adopted in the previous subsection. We do not plan to study the
high temperature aspects of this model here but focus on the
behaviour of the glueball mass and wave function in the deconfined
region.

The physical temperature $T=1/(aN_{t})$, is given via the lattice
parameters as follows:
\begin{equation}
T/\sqrt{\sigma} = \frac{1}{N_{t}\sqrt{K}}.
\label{eqn10}
\end{equation}
For completeness, we give a
temperature estimate of 1.125 in the units of string tension.
By setting the string tension to 420 MeV, we estimate a physical
temperature of $T\sim 1.25T_{c}$, where the $T_{c} \sim 360$ MeV at
pseudo-critical coupling $\beta_{c} = 1.87(2)$ \cite{Chernodub}.
One expects \cite{Svetitsky82} that the high temperature
phase has a massless photon  and the linear
potential is replaced by the two dimensional logarithmic Coulomb
potential. This logarithmic behaviour is equivalent to a power-law
dependence of the Wilson loop correlation function,
\begin{equation}
C({\bf r })= \langle P^{\dagger}({\bf r})P(0)\rangle \sim \mid
{\bf{r}}\mid^{-\eta (T)},
\label{eqn11}
\end{equation}
with an exponent which decreases as $T$ increases. Furthermore,
since the high-temperature phase of the gauge theory corresponds
to the ordered phase of the spin system, the predicted power-law
behaviour of the correlation function is just like that of a
two-dimensional U(1)-invariant spin system - a 2-D XY model.

\begin{figure}[!h]
\scalebox{0.45}{\includegraphics{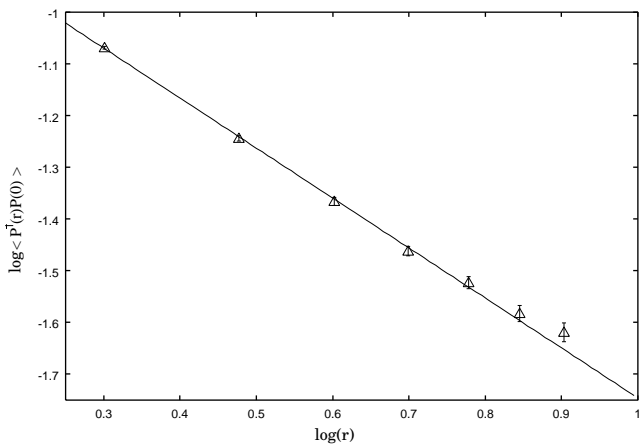}}
\caption{
\label{figcorzt}
The logarithmic plot of the correlation function  at $\beta =2.25$.
The straight line indicates power-law behaviour.}
\end{figure}

Fig. \ref{figcorzt} shows a plot  of correlation functions
versus separation. The straight line indicates the fit to the form
(\ref{eqn11}).
The finite  temperature phase transition is visible in the change
of the correlation function from exponential to power-law
behaviour. Thus it becomes evident that $T>T_{c}$ in our
simulation. It can be seen that form (\ref{eqn11}) fits the data
rather well. Nonetheless, our Monte Carlo simulations were unable
to confirm that the exponent is moving towards the value of 0.25
(that of the 2-D XY model \cite{Tobochnik79}) predicted for the
continuum theory. Our estimated value for the exponent is four
times larger than the predicted value. This indicates that our
$\beta$ value of 2.25 is not large enough to gives us reason to
hope that we are approaching continuum physics. An interesting
feature to explore in this context is whether the coupling to the
matter fields in the leading order ($\beta \rightarrow \infty$)
calculations will move the critical exponent towards the predicted
value.
\begin{figure}[!h]
\scalebox{0.45}{\includegraphics{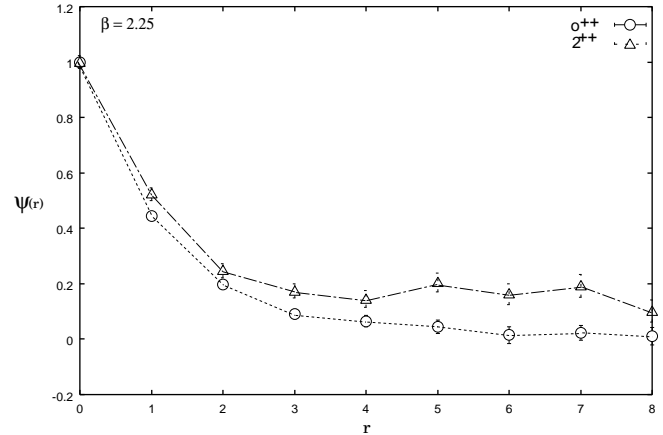}}
\caption{
\label{figzt}
Scalar and tensor glueball wave functions measured on $16^{2}\times 4$
lattice at $\beta = 2.25$.}
\end{figure}

In the deconfinement phase above the critical temperature,
glueballs are no more elementary excitations. At high temperatures
we have a plasma that behaves in bulk roughly like a free gas of
quarks and gluons, thus forming a new phase, i.e., the quark-gluon
plasma (QGP) phase. Above the critical temperature $T_{c}$,
properties such as confinement and chiral symmetry breaking
disappear. A detailed understanding of thermal glueballs gained
over the last decade can be found in
\cite{Ishii02,Detar87,Saumen98,Saumen03,Bialas04,Hart00} and the
references therein. As a result, quarks and gluons are liberated
and tremendous changes are expected in the mass spectrum. Fig.
\ref{figzt} shows the scalar and tensor wave functions obtained
through the same analysis as in Figs. \ref{fig0pp} and
\ref{fig2pp}. Our results indicate that no significant changes
occur in the scalar and tensor wave functions. Glueball masses
appear almost unaffected. By comparing the results at $T=0$ and
$T=360$, we observe an effective  mass reduction, ($am_{G}(T\sim
0)-am_{G}(T\sim 360)$), of about $4\%$, with statistical
uncertainties typically on less than a percent level, for $0^{++}$
and $2^{++}$ glueball modes. This appears to be a very small
change since we expect a rather continuous mass reduction of
glueballs in the deconfined phase. This might be due to the fact
that for zero momentum the power-law behaviour of the correlation
function leads at short distances to the spin-wave results, which
prevents us from seeing the massless excitations.

The non-vanishing effective masses  would suggest the presence of
glueball modes above $T_{c}$. Other work on finite temperature
SU(3) \cite{Ishii02,Detar87} has also confirmed the survival of
correlations above $T_{c}$ in the scalar and tensor colour-singlet
modes. However, these studies have shown that thermal mass changes
rather continuously across the critical temperature. The existence
of the effective mass gives rise to the possibility that some of
the nonperturbative effects survive in the deconfined phase, and
the colour-singlet modes exist as metastable states above $T_{c}$.
The metastable states in the ordered phase (large $\beta$) appear
to be caused by the unusually large separation of a vortex pair,
which may take many sweeps to recombine. Near the transition the
number of vortices increase, and some of them begin to unbind.
This eventually drives the system into a disordered phase as one
moves to the region $T<T_{c}$.

Whether bound or metastable modes, the glueballs can decay into
two or more gluons thus acquiring finite width which is expected
to become less negligible in the deconfined phase. Thus it becomes
important to take into account the effect of width in best-fit
analysis. This might also explain a very modest reduction of our
masses at $T>T_{c}$. However, from this study, it is not possible
to determine whether such colour singlet modes really survive
above $T_{c}$ as metastable states. An extensive systematic
analysis, of unquenched improved lattice QCD at finite
temperature, along these lines is under way \cite{Mushe06b}.

\section{Summary and Conclusion}
\label{sec4} We have studied wave functions and sizes of scalar
and tensor glueballs using improved 3-dimensional U(1) lattice
model. In this preliminary study we take a more direct approach to
the problem; instead of fixing a gauge or a path for the gluons,
we measure the correlation functions from our lattice operators
from spatially connected Wilson loops which, being the expectation
values of closed-loop paths, are gauge invariant. This approach
has the advantage that the disentangling of the glueball and
torelon is usually taken care of automatically by the choice of
Wilson or Polyakov loops. We observed that the  size of tensor
glueball is roughly two times larger than the size of the scalar
glueball. We believe that our estimates are more reliable than the
results obtained in Ref. \cite{Forcrand92}, where the size of the
tensor glueball was found to be $\sim 0.8$ fm,  four times as
large as its scalar counterpart. The predicted zero lattice
spacing results are not actually found by extrapolation to zero
lattice spacing, but are obtained instead from calculations at
$\beta$ of 2.2 of glueball size, with no accurate representation
of the effect of the absence of extrapolation. Also the results
were of limited interest because of their manifest dependence on
the gauge chosen and the problem of disentangling of the glueballs
and torelons.

Finally, for completeness, we extended our method to measure the
wave function and mass for a finite temperature deconfinement
phase. For the lowest $0^{++}$ and $2^{++}$ glueballs, no
significant mass reduction was observed in the deconfined phase,
while the wave functions remain almost unchanged. The existence of
the effective mass  indicates that colour-singlet modes may
survive in the deconfined phase as metastable states. In such a
case glueball decay and decay width, as spectral component, in the
deconfinement phase are the most feasible candidates for a more
reliable analysis for the future studies.

\begin{acknowledgments}
We are grateful to  D. Leinweber and C. Hamer for a number of
valuable suggestions which provided the impetus for much of this
work. We are also grateful for access to 128 node DeepSuper -21C
computing facility at the Shenzhen University. This work was
supported by the Guangdong Provincial Ministry of Education and
Jinan University.
\end{acknowledgments}

\end{document}